\documentclass{article}

\usepackage[preprint]{corl_2026} 

\usepackage{amsmath,amssymb,amsfonts}
\usepackage{graphicx}
\usepackage{float}
\usepackage{booktabs}
\usepackage{multirow}
\usepackage{xcolor}
\usepackage{colortbl}

\newcommand{\runin}[1]{\par\noindent\textbf{#1.}\ \ignorespaces}

\title{\textbf{SEAOTTER}: \textbf{S}ensor \textbf{E}mbedded \textbf{A}utoencoding with \textbf{O}ne-\textbf{T}ime \textbf{T}ranscode for \textbf{E}fficient \textbf{R}econstruction}

\author{
  Dan Jacobellis and Neeraja J. Yadwadkar\\
  Department of Electrical and Computer Engineering\\
  The University of Texas at Austin\\
  Austin, TX 78712, USA\\
  \texttt{danjacobellis@utexas.edu}, \texttt{neeraja@austin.utexas.edu}
}

\begin{document}
\maketitle


\begin{abstract}
In robotics systems, vast amounts of visual data are easily captured at high resolution using low-cost, low-power hardware.
Yet, limited bandwidth and on-device compute resources prevent full utilization when transmitted via conventional codecs like JPEG/MPEG.
Newer codecs, like AV1/AVIF, improve the rate-distortion trade-off, but demand far more resources for encoding, impractical without custom ASICs.
Recent asymmetric autoencoders deliver high quality under extreme power and bandwidth constraints, but add prohibitive decoding cost and use bespoke formats that ignore decades of infrastructure built around standards like JPEG.
To address these limitations, we introduce a compression framework for cloud robotics based on a \textbf{S}ensor \textbf{E}mbedded \textbf{A}utoencoder paired with a \textbf{O}ne-\textbf{T}ime \textbf{T}ranscode for \textbf{E}fficient \textbf{R}econstruction (SEAOTTER).
Because the sensor, cloud, and consumer stages face very different power and bandwidth budgets, SEAOTTER combines the compactness of a learned latent with the broad usability of a standard JPEG file.
Since naive transcoding degrades performance, we propose a learnable JPEG color and quantization transform that enables increased accuracy for global, dense, and vision-language-based perception.
Using SEAOTTER, we train both general-purpose and task-aware transcoding pipelines for a pre-trained, frozen encoder. At a compression ratio of 200:1 and compared to AVIF, we observe $7\times$ faster encoding, $3.5\times$ faster decoding, and +8\% ImageNet top-1 accuracy, while retaining compatibility with JPEG infrastructure. Our code is available at \url{https://github.com/UT-SysML/seaotter}.

\end{abstract}

\keywords{Cloud robotics, Representation learning, Image compression}


\begin{figure}[H]
\centering
\includegraphics[width=\textwidth]{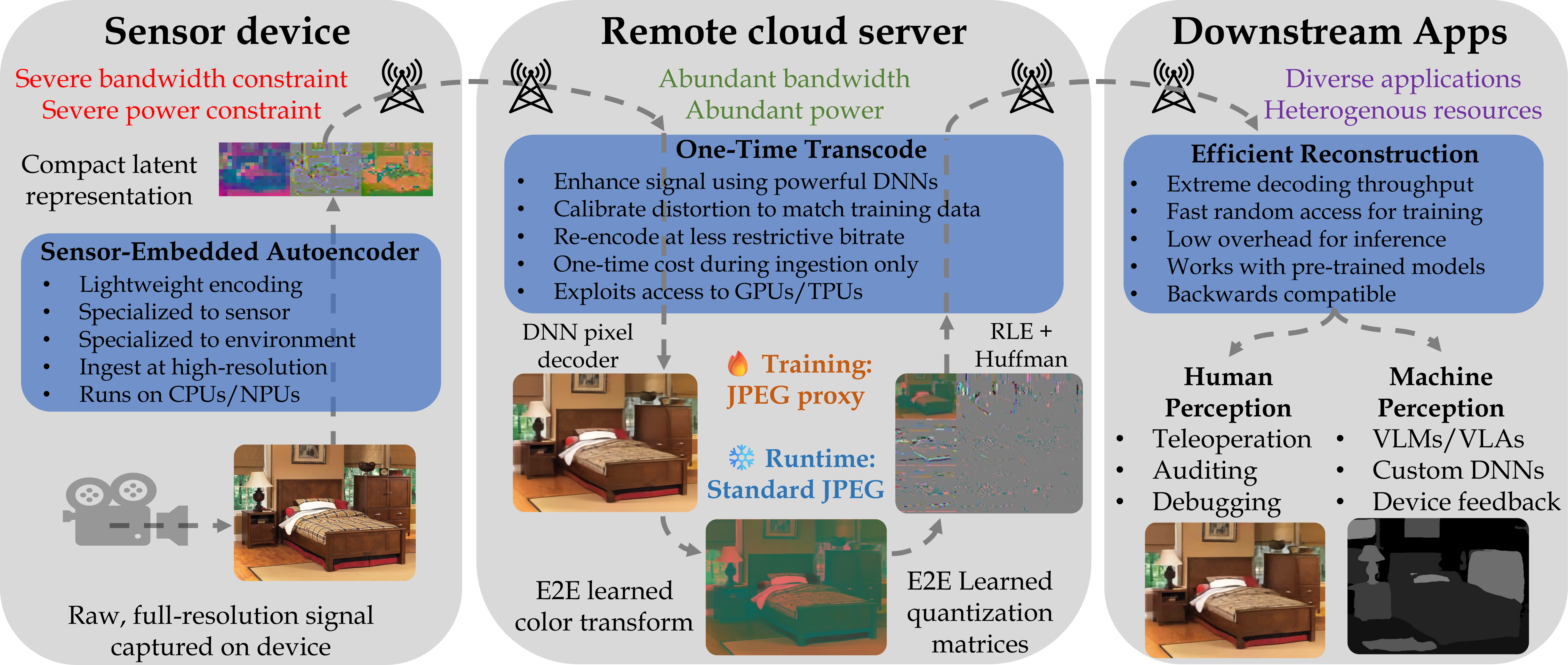}
\caption{Overview of SEAOTTER design and workflow.}
\label{fig:system}
\end{figure}

\section{Introduction}
\label{sec:intro}

The staggering economic scale of the smartphone market has driven extraordinary advances in image sensing: modern low-cost, low-power sensors let small, battery-powered robots and wearables capture billions of pixels per second at a fidelity once reserved for earth-observation satellites, consuming on the order of $10^{-11}$~joules per pixel~\citep{chen201412, kim2022fully}.
Yet fully utilizing these information-dense signals on-device is prohibitive, because the most capable ViT- and CNN-based perception systems require FLOPs that scale super-linearly with resolution~\citep{beyer2024vitspeed}; it is common to instead use only low-resolution feeds and discard the rest.
Cloud-robotics approaches---remote inference~\citep{jacobellis2025dedelayed}, split inference~\citep{kang2017neurosurgeon,matsubara2022split}, and collaborative inference~\citep{wang2024end,gao2025feature}---offload this computation to datacenters where power is abundant, but on-device power and bandwidth then demand extreme compression to reach the cloud.
For example, a 1080p 30~fps stream over a 25~Mbps Wi-Fi channel requires a compression ratio of about 60:1, and a 480p stream over a 1~Mbps BLE channel about 288:1~\citep{balasubramanian2009energy,carroll2010analysis,gupta20243,tosi2017performance}.
Conventional codecs like JPEG/MPEG meet these ratios only at severe perceptual cost~\citep{jacobellis2024machine}; newer standards (AV1/AVIF) and decoding-efficient asymmetric autoencoders (DE-AAEs)~\citep{yang2023computationally} improve the rate--distortion trade-off but demand prohibitively expensive encoding.
Encoding-efficient asymmetric autoencoders (EE-AAEs)~\citep{hojjat2025mcucoder,jacobellis2025learned,jacobellis2026liveaction} invert this trade-off, pairing a lightweight encoder with an expensive DNN decoder that removes the artifacts of severe dimension reduction.
We build on FRAPPE~\citep{jacobellis2026frappe}, whose encoder costs only 10--100 MAC/pixel---at low bitrates, less than JPEG.

These EE-AAEs, however, are impractical on the consumer side: their DNN decoders are costly to run, and their bespoke latents are incompatible with the decades of infrastructure built around JPEG---ML frameworks, fast dataloaders, web browsers, and hardware codecs baked into ASICs and SoCs.
Decoding cost is especially punishing under the encode-once, decode-many lifecycle of modern workloads: a training run re-reads each file once per epoch with fresh augmentations, so any per-decode overhead is multiplied by the consumption count.
A single up-front transcode into a cheaper-to-decode artifact is therefore favorable whenever a file is read more than once---which is why JPEG/M-JPEG remains ubiquitous across robotics.

To address these limitations, we introduce SEAOTTER (Fig.~\ref{fig:system}), a \textbf{S}ensor \textbf{E}mbedded \textbf{A}utoencoder with a \textbf{O}ne-\textbf{T}ime \textbf{T}ranscode for \textbf{E}fficient \textbf{R}econstruction that reconciles resource-constrained sensors with data-hungry consumers through three goals detailed in Section~\ref{sec:method}: high-throughput sensor-side encoding, end-to-end task adaptability, and universal consumer-side compatibility.

\begin{figure}[!t]
\centering
\includegraphics[width=\textwidth]{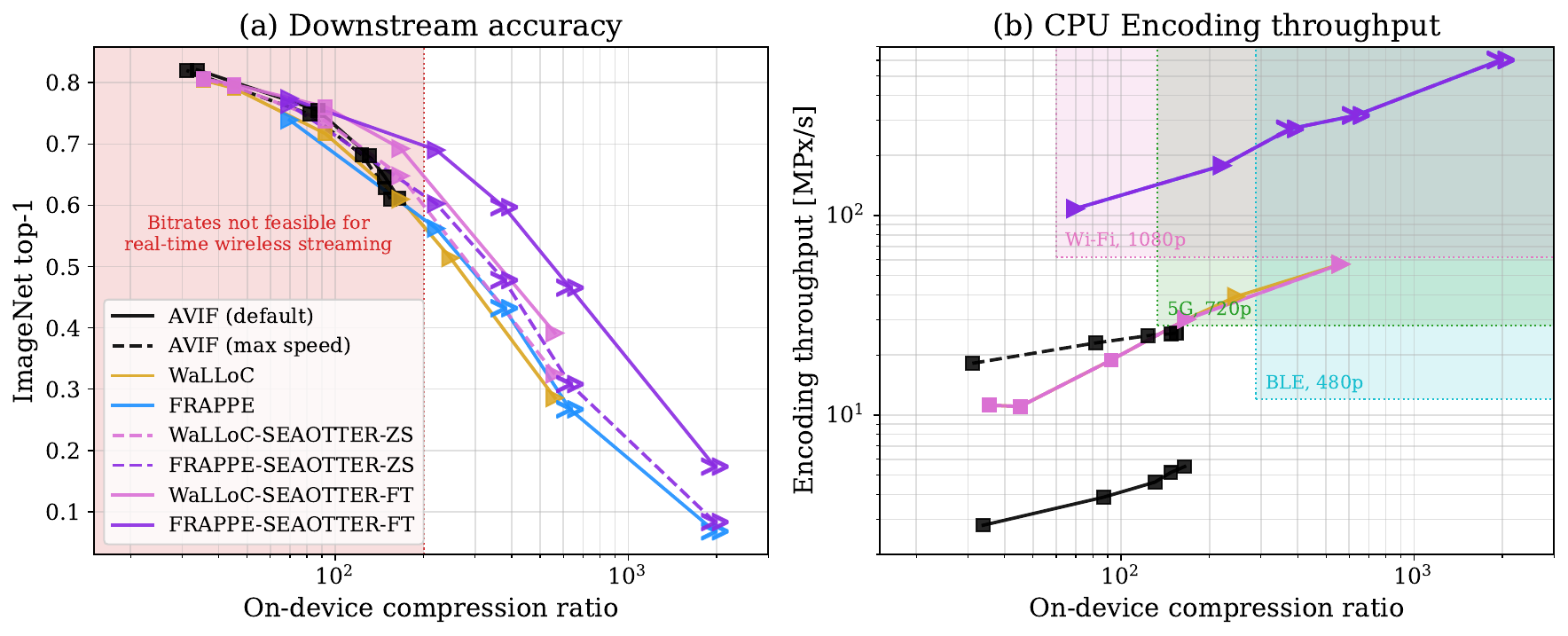}
\caption{(a) Classification accuracy and (b) CPU encoding throughput vs.\ on-device compression ratio. Shaded regions mark the compression ratio and throughput needed for 1080p/30 over Wi-Fi (25 Mbps), 720p/30 over 5G (5 Mbps), and 480p/30 over BLE (1 Mbps). $\blacksquare$, $>$, and $\gg$ mark poor, fair, and excellent on-device suitability; configurations in the red region are poorly suited.}
\label{fig:main_results}
\end{figure}

\paragraph{Sensor-embedded encoding under extreme resource constraints.}
Robotic, wearable, and remote-sensing platforms run their image sensors against strict power, thermal, and uplink-bandwidth budgets, so the sensor-side encoder must spend orders of magnitude less compute per pixel than a hyperprior~\citep{balle2018variational,minnen2018joint}, a vanilla JPEG encoder, or modern codecs like AVIF, whose run-time rate-distortion optimization and multi-stage in-loop filtering reach a per-pixel cost that production deployments meet only with dedicated hardware encoders~\citep{bossen2021vvc}. SEAOTTER instead uses an EE-AAE~\citep{hojjat2025mcucoder,jacobellis2025learned,jacobellis2026liveaction} built on a pre-trained FRAPPE codec~\citep{jacobellis2026frappe}, chosen for its high encoding efficiency and low-overhead variable-rate and progressive coding---crucial for systems operating under fluctuating bandwidth and shared CPU/NPU load.

\paragraph{Learning specialized representations via end-to-end optimization.}
To support diverse robotics applications, the codec must handle arbitrary sensors and conditions---high motion, aerial views, poor lighting, fish-eye distortion. JPEG uses fixed color transforms and quantization matrices tuned to human perception; SEAOTTER instead learns these from data while staying compatible with standard JPEG hardware and software, specializing to the camera, environment, and downstream model. We freeze the FRAPPE encoder and fine-tune the FRAPPE decoder and JPEG color/quantization parameters; jointly optimizing the encoder could yield further gains.

\paragraph{Flexible and efficient decoding.}
The cloud-side transcode produces standard-compliant JPEG files with the custom quantization matrices embedded in metadata. The standard RGB-YUV color transform is forgone, and the codec is sandwiched between a learned color transform and a companding nonlinearity that enforces the limited dynamic range. For bespoke machine-vision applications, decoding is then \emph{faster} than standard JPEG, since the inverse color transform can be skipped to operate directly in the learned color space~\citep{gueguen2018faster,ehrlich2019deep}. For pre-trained models that accept sRGB inputs and cannot be fine-tuned (e.g., billion-parameter foundation models, VLMs, and VLAs), the only overhead is a single post-filter (${\sim}81$ MACs/pixel).

\paragraph{Contributions.}
Using SEAOTTER, we (i) frame cloud-robotics compression as a three-way sensor / cloud / consumer asymmetry under an encode-once, decode-many lifecycle; (ii) introduce an end-to-end learned JPEG codec---color transform, quantization, and rate proxy trained de novo---that beats the ITU T.81 tables; and (iii) show across global, dense, and vision-language tasks that the one-time transcode \emph{increases} downstream accuracy over the underlying autoencoder while emitting standard JPEG files (Fig.~\ref{fig:main_results}).




\section{Proposed method: design and implementation}
\label{sec:method}

\runin{Overview and workflow}
SEAOTTER's pipeline has three stages separated by two compressed bitstreams: a frozen sensor-embedded analysis transform produces a quantized $\text{int}8$ latent that is losslessly compressed and transmitted over the wireless uplink; at the cloud, a heavy synthesis transform reconstructs an intermediate pixel image, which an end-to-end learned JPEG codec re-encodes as a standard JPEG file---a transcode paid exactly once per captured frame. The on-disk artifact is thereafter a plain JPEG file, decoded by every downstream consumer with a vanilla JPEG decode followed by a single learned inverse color transform. Fig.~\ref{fig:system} overviews the workflow, described next.

Let $x\in\mathbb{R}^{3\times H\times W}$ denote an input RGB image, normalized to $[-1,1]$. SEAOTTER composes a sensor-side analysis transform $\mathcal{G}_{\!A}$, a lossless transmission channel $\mathcal{C}$, a cloud-side synthesis transform $\mathcal{G}_{\!S}$, a learned color transform $\mathcal{F}$ with inverse $\mathcal{F}^{-1}$, and a JPEG codec $\mathcal{J}_Q$ parameterized by a learned quantization matrix $Q$:
\begin{equation}
  \hat{x} \;=\; \mathcal{F}^{-1} \,\circ\, \mathcal{J}_Q \,\circ\, \mathcal{F} \,\circ\, \mathcal{G}_{\!S} \,\circ\, \mathcal{C} \,\circ\, \mathcal{G}_{\!A}(x).
  \label{eq:pipeline}
\end{equation}
Here $\mathcal{G}_{\!A}$ is the frozen FRAPPE encoder and $\mathcal{G}_{\!S}$ the matching FRAPPE decoder (fine-tuned below); the lossless channel $\mathcal{C}$ packages JPEG-LS~\citep{weinberger2000loco} entropy coding, uplink transmission, and cloud-side decoding, so for Eq.~\eqref{eq:pipeline} it is the identity; $(\mathcal{F}, \mathcal{F}^{-1})$ is the invertible learned color transform; and $\mathcal{J}_Q$ is the single inherently lossy step, a standard JPEG encode--decode round-trip with the learned quantization matrix $Q$.

\runin{FRAPPE encoder for variable rate compression under extreme resource constraints}
$\mathcal{G}_{\!A}$ is a FRAPPE~\citep{jacobellis2026frappe} encoder, which projects input patches of varying scales (from $32{\times}32$ to $4{\times}4$) to scalar values. Its cost is dominated by the linear projections and amounts to roughly $10$--$100$ MAC/pixel depending on the operating point---two orders of magnitude lower than the smallest learned hyperprior codecs~\citep{balle2018variational,minnen2018joint}. FRAPPE's residual training procedure sorts the latent channels in coarse-to-fine order, so a single set of encoder weights serves every supported rate point ($n\in\{3,6,9,12,15\}$): the sensor selects its operating point by transmitting a prefix of the channels rather than re-encoding. We freeze $\mathcal{G}_{\!A}$ throughout, matching the asymmetric-capacity stance of WaLLoC~\citep{jacobellis2025learned} and LiVeAction~\citep{jacobellis2026liveaction}, where sensor-side compute is a hard budget rather than a tunable axis. After encoding, the int8 latents are losslessly compressed (the framework is agnostic to the specific lossless codec).

\runin{Fine-tuned FRAPPE decoder for application-specific signal enhancement and calibration}
The cloud-side synthesis transform $\mathcal{G}_{\!S}$ is the FRAPPE decoder ($\sim$$57\text{M}$ parameters). Unlike $\mathcal{G}_{\!A}$, $\mathcal{G}_{\!S}$ is \emph{fine-tuned} against the downstream task loss with the encoder still frozen; its RGB output drifts toward a distribution that a JPEG-pretrained consumer-side backbone (Section~\ref{sec:eval}) reads more accurately. Because $\mathcal{G}_{\!A}$ is frozen, every fine-tuned $\mathcal{G}_{\!S}$ snapshot is interchangeable at runtime: the same transmitted latent decodes to different RGB outputs depending on the chosen snapshot, so a single uplink stream can serve multiple downstream tasks simultaneously. The fine-tune deliberately sacrifices pixel-domain reconstruction PSNR for higher downstream accuracy after the transcode, specializing $\mathcal{G}_{\!S}$'s output for the JPEG step that follows.

\begin{figure}[!t]
\centering
\includegraphics[width=\textwidth]{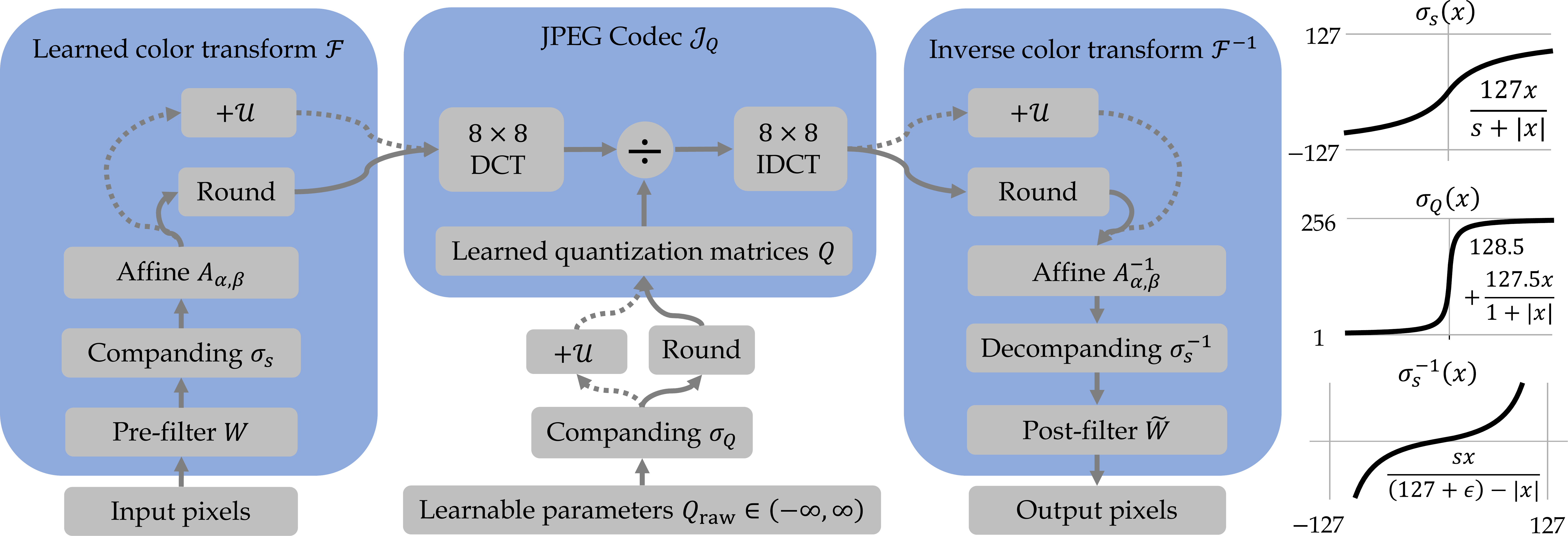}
\caption{JPEG workflow with learned color transform and quantization and visualization of companding/decompanding functions. Dotted lines indicate the signal path during training.}
\label{fig:jpeg_sandwich}
\end{figure}

\runin{JPEG sandwich}
The cloud-side decoder's RGB output enters a learned JPEG sandwich: a forward color transform $\mathcal{F}$ into a JPEG-friendly representation, a standard JPEG encode with a learned $3{\times}8{\times}8$ quantization matrix $Q$, and an inverse color transform $\mathcal{F}^{-1}$ at the consumer. The closest prior art is the sandwiched codec of \citet{guleryuz2021sandwiched,guleryuz2024sandwiched}, which wraps U-Net pre- and post-processors around a standard codec, trained end-to-end on a per-image rate--distortion proxy. SEAOTTER differs in three ways: (i) its color transform $\mathcal{F}$ is a lightweight $3{\times}3$ convolution plus companding rather than a U-Net, so the consumer-side decode pays at most a vanilla JPEG-decode cost plus a few thousand floating-point operations per pixel; (ii) it is trained \emph{de novo} with no codec warm-starts, so its win over standard JPEG comes from representation rather than bookkeeping; and (iii) a single learned $(\mathcal{F}, \mathcal{F}^{-1})$ pair is shared across all $K$ rate points, with $K$ independent quantization matrices $Q^{(1)},\dots,Q^{(K)}$ specializing the per-rate behavior. Fig.~\ref{fig:jpeg_sandwich} shows the resulting workflow.

\runin{De novo learnable wrapper filter and color transform}
$\mathcal{F}$ composes three operators: a $3{\times}3$ wrapper filter $\mathrm{Conv}_W$ with learnable kernel $W$---a full $3$-input, $3$-output convolution that jointly filters spatially and mixes the three RGB channels, so it is this operator (not the per-channel stages that follow) that realizes the learned color space---a per-channel softsign companding $\sigma_s$ with learnable scale $s\in\mathbb{R}^3_+$ that confines each channel to the signed 8-bit range $(-127, 127)$, and a per-channel affine $A_{\alpha,\beta}$ with learnable scale $\alpha\in\mathbb{R}^3$ and offset $\beta\in\mathbb{R}^3$ that packs the result into the unsigned 8-bit range $[0, 255]$:
\begin{align}
  \mathcal{F}(x) &\;=\; A_{\alpha, \beta} \,\circ\, \sigma_s \,\circ\, \mathrm{Conv}_W(x), \label{eq:F}\\
  \mathcal{F}^{-1}(y) &\;=\; \mathrm{Conv}_{\widetilde W} \,\circ\, \sigma_s^{-1} \,\circ\, A_{\alpha, \beta}^{-1}(y), \label{eq:Finv}
\end{align}
where $\mathcal{F}^{-1}$ mirrors $\mathcal{F}$ but with an \emph{independently-learned} wrapper-filter kernel $\widetilde W$ (a $3{\times}3$ convolution is not in general algebraically invertible). $A_{\alpha, \beta}^{-1}$ and $\sigma_s^{-1}$ are the closed-form algebraic inverses of the corresponding forward operators, with $\alpha, \beta, s$ shared between $\mathcal{F}$ and $\mathcal{F}^{-1}$. The unit-step rounding at the JPEG codec's boundaries is handled by the canonical three-mode contract of the hyperprior codecs~\citep{balle2017end,balle2018variational,minnen2018joint}: during training it is replaced with independent additive uniform noise $u\sim\mathcal{U}[-\tfrac{1}{2},\tfrac{1}{2}]$, during evaluation it uses the continuous output of the preceding operator, and at deployment an explicit $\mathrm{round}(\cdot)$ is applied outside the forward pass. The softsign companding inside $\mathcal{F}$ confines its pre-quantization output to the 8-bit range regardless of input magnitude, so the contract holds for arbitrary pixel-domain dynamic ranges without per-sensor calibration.

All learnable parameters of $\mathcal{F}$ and $\mathcal{F}^{-1}$---the wrapper-filter kernels $W$ and $\widetilde W$, the softsign scales $s$, and the affine $(\alpha, \beta)$---are initialized so that the composed map is approximately the algebraic identity at step zero, \emph{not} the JFIF $\text{RGB}{\to}\text{YCbCr}$ matrix: warm-starting from JFIF would have the network \emph{deviate} from the codec we are trying to displace rather than \emph{discover} a color transform, so with identity initialization the only inductive bias is the architectural shape and everything chromatic falls out of the rate--distortion loss on data. At inference time, $\mathcal{F}$'s three-channel output is written byte-for-byte into the JPEG file with \texttt{subsampling=0} (true $4{:}4{:}4$); since the channels are not chroma in the conventional sense, the JPEG decoder must skip the standard $\text{YCbCr}{\to}\text{RGB}$ color conversion. Both options---\texttt{subsampling=0} and the skipped color conversion---are standard settings exposed by any compliant JPEG implementation, so the on-disk artifact remains decodable by any standards-compliant codec. We verified that, with identity weights, this gives a bit-exact $\text{RGB}{\leftrightarrow}\text{RGB}$ round-trip.

\runin{De novo learnable DCT-domain quantization matrices}
For each rate point $k=1,\dots,K$, an unconstrained $3{\times}8{\times}8$ parameter tensor $Q^{(k)}_{\text{raw}}$ maps to a JPEG quantization matrix in the open range $(1,256)$ via a softsign-plus-affine reparameterization,
\begin{equation}
  Q^{(k)} \;=\; 128.5 \;+\; 127.5 \,\cdot\, \operatorname{softsign}\!\bigl(Q^{(k)}_{\text{raw}}\bigr).
  \label{eq:qtable}
\end{equation}
During training, $Q^{(k)}$ is the continuous divisor of the $8{\times}8$ block DCT; at deployment it is rounded and clamped to integers in $[1,255]$. The softsign-plus-affine parameterization is borrowed from FRAPPE's encoder companding: it keeps the gradient finite as $Q^{(k)}$ approaches either boundary of the JPEG-legal range. Per-rate independence lets each $Q^{(k)}$ specialize to its operating point, while the shared $(\mathcal{F}, \mathcal{F}^{-1})$ keeps sensor- and consumer-side costs constant across rates. The learned $Q^{(k)}$ matrices and the resulting (approximately YCgCo) color space are visualized in Appendix~\ref{app:quant}.

\runin{Improved JPEG rate proxy}
End-to-end training requires a differentiable proxy for the JPEG file size the codec actually produces. We use a sparsity-aware run-length surrogate that models JPEG's zigzag AC coding: per $8{\times}8$ block, the bit count is a smooth nonzero gate $\tanh(c_k^2)$ times $\log_2\!\bigl(1+|c_k|/Q^{(k)}_k\bigr)$ plus a fixed Huffman-overhead constant, summed over all blocks. This is closely related to the $\log\!\bigl(1+|x_i|/\Delta\bigr)$ proxy of \citet{guleryuz2021sandwiched}, with two changes: the soft gate $\tanh(c_k^2)$ captures the dominant cost of an AC block---whether its coefficients fall within the zero run---and the per-block overhead constant absorbs the Huffman-table bits the bitstream-level proxies omit. A single per-rate scalar $\alpha^{(k)}$, fit on held-out images, calibrates the surrogate to the real JPEG bits-per-pixel of a standards-compliant encoder; we denote the calibrated proxy $\mathrm{bpp}^{(k)}(x, Q^{(k)})$.

\runin{De novo training of JPEG color and quantization}
We train the shared color-transform pair $(\mathcal{F}, \mathcal{F}^{-1})$ and the $K$ rate-specific quantization matrices $Q^{(1)},\dots,Q^{(K)}$ jointly, end-to-end, against a multi-rate rate--distortion objective:
\begin{equation}
  \mathcal{L}_{\text{total}}
  \;=\;
  \sum_{k=1}^{K} w_k \cdot
    \Bigl[\,\log_{10} \mathrm{MSE}_k(x, \hat{x}_k) \;+\; \lambda_k \cdot \mathrm{bpp}^{(k)}\!\bigl(x, Q^{(k)}\bigr)\,\Bigr],
  \label{eq:loss}
\end{equation}
where $\hat{x}_k$ is the reconstruction at rate point $k$ under the shared $(\mathcal{F}, \mathcal{F}^{-1})$ pair and the rate-specific $Q^{(k)}$, $\mathrm{bpp}^{(k)}$ is the calibrated rate proxy, $\lambda_k>0$ is the Lagrange multiplier trading rate against distortion, and $w_k>0$ is a per-rate loss weight that arbitrates between rate paths when one would otherwise dominate the gradient. The shared pair is updated by gradients from all $K$ terms of Eq.~\eqref{eq:loss} simultaneously, so the color transform learns a rate-agnostic representation; each $Q^{(k)}$ receives gradient only from its own term, so the quantization matrices specialize to their operating points.

The codec is trained \emph{de novo}: $(\mathcal{F}, \mathcal{F}^{-1})$ is initialized to the algebraic identity (no JFIF warm-start), each $Q^{(k)}$ is initialized from random Gaussian noise (no \texttt{quality=Q} warm-start), and the JPEG standard's precomputed Huffman tables are used at runtime instead of per-image optimized tables. Full training details, including the headline $K{=}3$ rate weights $(\lambda_k, w_k)$, are in Appendix~\ref{app:training_recipe}.

\runin{Optional inverse color transform and post-filter}
The consumer-side decode is a vanilla JPEG decode followed by $\mathcal{F}^{-1}$, which recovers the displayable RGB from the three-channel $\text{uint}8$ output. Crucially, $\mathcal{F}^{-1}$ is \emph{optional}: downstream applications that train or fine-tune their own consumer-side model can skip it and operate directly on the JPEG-decoded coefficients, analogous to prior work on JPEG-domain learning systems that consume YUV (or YCoCg) directly~\citep{gueguen2018faster,ehrlich2019deep}---the skipped inverse-conv is absorbed into the first layer of the downstream model with no loss of expressivity. SEAOTTER therefore coexists with both legacy JPEG-consuming pipelines (which apply $\mathcal{F}^{-1}$) and JPEG-domain learning pipelines (which skip it).


\section{Performance evaluation}
\label{sec:eval}

We evaluate SEAOTTER in terms of the rate--distortion--complexity trade-off. Rate is measured in bits per pixel (bpp), reported as both the transmission rate (tbpp, uploaded from the sensor) and the storage rate (sbpp, the transcoded JPEG file); the compression ratio is $\mathrm{CR} = 24/\mathrm{bpp}$. Distortion is measured via standard metrics (PSNR, SSIM~\citep{wang2004ssim}, LPIPS~\citep{zhang2018lpips}, DISTS~\citep{ding2020dists}) and downstream task accuracy, and complexity via on-device CPU encoding throughput (megapixels per second). We compare against AVIF, WaLLoC, and FRAPPE, and evaluate both a zero-shot SEAOTTER pipeline (pre-trained for MSE on a general-purpose dataset~\citep{li2023lsdir}) and a task-specific fine-tuned pipeline. We additionally evaluate the learned JPEG codec as a standalone system against the standard ITU T.81~\citep{itu1992t81} colorspace and quantization tables, with and without chroma subsampling.

\runin{Models, datasets, and task-specific performance metrics}
We evaluate downstream task accuracy on three tasks chosen to span global, dense, and VLM-style inference. For \textbf{global classification} (cls), we use ImageNet val ($50{,}000$ images)~\citep{deng2009imagenet} with a ConvNeXt-Tiny teacher~\citep{liu2022convnet}, reporting top-1/top-5 accuracy. For \textbf{dense prediction} (seg), we use ADE20K val ($2{,}000$ images)~\citep{zhou2017scene} with a UperNet-ConvNeXt-Tiny teacher~\citep{xiao2018unified}, reporting mIoU. For \textbf{VLM/VLA-style zero-shot prediction} (clip), we use ImageNet val with the SigLIP-2 base-patch16-naflex encoder~\citep{tschannen2025siglip2}, reporting zero-shot top-1. Preprocessing is squash to the task resolution (cls $384^2$, seg $512^2$) or naflex (clip), which also sets the bpp denominator. Codec baselines are AVIF (default and max-speed, $\texttt{s10}$), FRAPPE, and WaLLoC; SEAOTTER variants are denoted \mbox{SEAOTTER-ZS} (zero-shot sandwich) and \mbox{SEAOTTER-FT} (decoder + sandwich fine-tuned for the target task). Standalone-codec baselines use ITU T.81 with and without chroma subsampling (Appendix~\ref{app:codec_kodak}). Teacher checkpoint IDs, naflex hyperparameters, the per-task no-codec accuracy ceilings, and timing hardware are reported in Appendix~\ref{app:experiment_details}.

\begin{figure}[!t]
\centering
\includegraphics[width=0.9\textwidth]{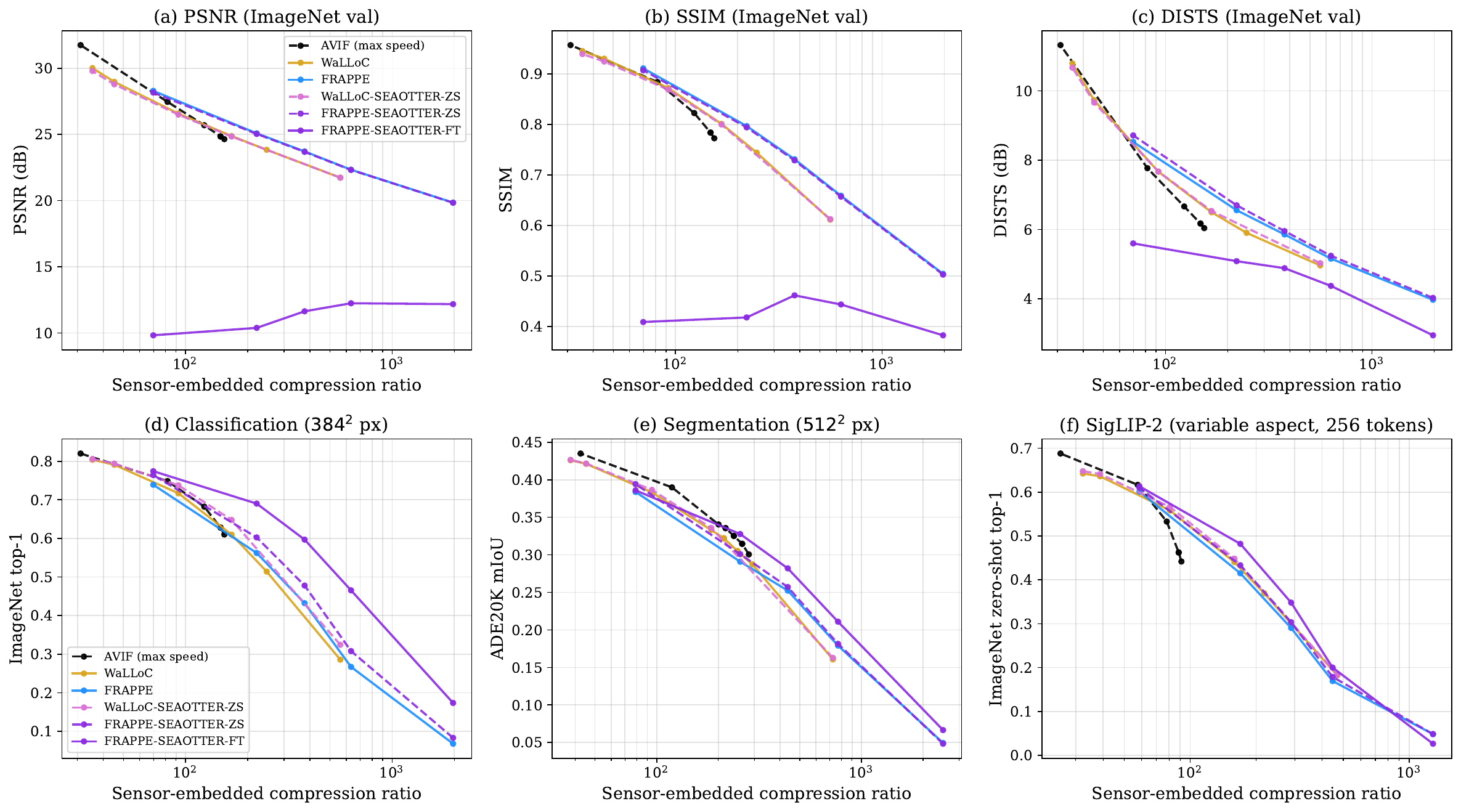}
\caption{Rate--distortion--accuracy trade-offs vs.\ transmit compression ratio: (top) SSIM and DISTS; (bottom) ADE20K mIoU and SigLIP-2 zero-shot top-1. SEAOTTER-ZS leads the perceptual-quality axes.}
\label{fig:rd_metric_panels}
\end{figure}

Figure~\ref{fig:main_results} summarizes the rate--accuracy--throughput trade-off of SEAOTTER variants against AVIF, WaLLoC, and FRAPPE; Figure~\ref{fig:rd_metric_panels} adds reconstruction-quality axes and Figure~\ref{fig:rd_storage} (Appendix~\ref{app:storage_rd}) the storage-rate axis; Appendix~\ref{app:summary} (Fig.~\ref{fig:radar}) gives a multi-axis overview. Table~\ref{tab:headline} groups three task subsections (cls, seg, clip) and within each reports every pipeline at a single per-task matched-rate operating point: FRAPPE / SEAOTTER variants stay at $n{=}12$, and each conventional baseline (AVIF, AVIF max-speed, WaLLoC) uses the lowest-bpp op still strictly above FRAPPE $n{=}12$ on that dataset (recomputed per task).

\begin{table}[!t]
\centering
\small
\caption{Summary of machine perception performance for images compressed at roughly 1--3~kB.}
\label{tab:headline}
\resizebox{\textwidth}{!}{\begin{tabular}{lrrrrrr}
\toprule
 & AVIF & AVIF (max-speed) & FRAPPE & WaLLoC & SEAOTTER-ZS & SEAOTTER-FT \\
\midrule
\multicolumn{7}{l}{\textit{ImageNet-1k ($384^2$)}} \\[2pt]
Operating point & $q=1$ & $q=1$ & $n=12$ & $p=16$ & $n=12$ & $n=12$ \\
Transmit CR & 165 & 154 & \textbf{221} & 167 & \textbf{221} & \textbf{221} \\
Storage CR & 165 & 154 & \textbf{221} & 167 & 19 & 27 \\
Top-1 Accuracy (\%) & 61.15 & 61.02 & 56.22 & 60.98 & 60.25 & \textbf{69.02} \\
Encode (MPx/s) & 5.51 & 25.73 & \textbf{177.76} & 30.17 & \textbf{177.76} & \textbf{177.76} \\
Decode (MPx/s) & 19.75 & 19.53 & 0.68 & 3.94 & 65.35 & \textbf{67.97} \\
\midrule
\multicolumn{7}{l}{\textit{ADE20k ($512^2$)}} \\[2pt]
Operating point & $q=5$ & $q=6$ & $n=12$ & $p=11$ & $n=12$ & $n=12$ \\
Transmit CR & \textbf{279} & 238 & 256 & 248 & 256 & 256 \\
Storage CR & \textbf{279} & 238 & 256 & 248 & 20 & 46 \\
mIoU (\%) & 32.75 & 32.51 & 29.09 & 30.51 & 30.09 & \textbf{32.77} \\
Encode (MPx/s) & 5.43 & 32.77 & \textbf{256.37} & 40.31 & \textbf{256.37} & \textbf{256.37} \\
Decode (MPx/s) & 19.79 & 19.55 & 0.68 & 3.24 & 65.35 & \textbf{67.97} \\
\midrule
\multicolumn{7}{l}{\textit{ImageNet-1k (naflex)}} \\[2pt]
Operating point & $q=1$ & $q=1$ & $n=12$ & $p=16$ & $n=12$ & $n=12$ \\
Transmit CR & 96 & 91 & \textbf{169} & 159 & \textbf{169} & \textbf{169} \\
Storage CR & 96 & 91 & \textbf{169} & 159 & 16 & 37 \\
SigLIP Accuracy (\%) & 42.59 & 44.19 & 41.51 & 44.05 & 43.34 & \textbf{48.22} \\
Encode (MPx/s) & 3.79 & 19.45 & \textbf{96.74} & 19.51 & \textbf{96.74} & \textbf{96.74} \\
Decode (MPx/s) & 19.75 & 19.53 & 0.68 & 3.94 & 65.35 & \textbf{67.97} \\
\bottomrule
\end{tabular}
}
\end{table}

\runin{Transcode increases downstream accuracy}
At matched transmit-bpp ($0.109$, $\text{CR}{=}221{:}1$), \mbox{SEAOTTER-FT} achieves $69.02\%$ ImageNet top-1 versus $56.22\%$ for FRAPPE alone---a $+12.80$ pp margin (Table~\ref{tab:headline}). The gap widens at lower bitrates: at $n{=}6$ (transmit-bpp $0.038$), \mbox{SEAOTTER-FT} reaches $46.55\%$ where FRAPPE-only gives $26.70\%$, a $+19.85$ pp improvement. Even the zero-shot variant (\mbox{SEAOTTER-ZS}, no task-aware fine-tune) recovers $+4.03$ pp over FRAPPE at $n{=}12$. The same effect appears on ADE20K segmentation (mIoU $+3.68$ pp for \mbox{SEAOTTER-FT} over FRAPPE at $n{=}12$) and SigLIP-2 zero-shot classification (top-1 $+6.71$ pp).

\runin{Pareto dominance on machine-perception tasks}
Under the matched-rate selection (Table~\ref{tab:headline}; Fig.~\ref{fig:rd_metric_panels}), \mbox{SEAOTTER-FT} leads ImageNet top-1 by $+7.87$ pp over AVIF and $+8.00$ pp over \mbox{AVIF-max-speed}, and leads SigLIP-2 zero-shot top-1 by $+5.63$ pp and $+4.03$ pp respectively---despite both baselines spending more bits per pixel. On ADE20K segmentation---the one axis where conventional baselines previously led at matched rate---\mbox{SEAOTTER-FT} now ties for first place ($+0.02$ pp over AVIF, $+0.26$ pp over AVIF-max-speed, $+3.68$ pp over FRAPPE alone).

\runin{Storage-side compression ratio}
Figure~\ref{fig:rd_storage} (Appendix~\ref{app:storage_rd}) re-plots the three task accuracies against the \emph{storage} compression ratio of the on-disk artifact. Against its architecturally-fair reference (FRAPPE followed by a vanilla ITU T.81 transcode at the same transmit-bpp), the \mbox{SEAOTTER-FT} artifact at $n{=}12$ is $13.7\%$ smaller \emph{and} yields $+8.19$ pp higher ImageNet top-1.

\runin{Sensor-side encoding throughput}
All SEAOTTER variants inherit the same frozen FRAPPE encoder, so the sensor-side budget is identical to the FRAPPE-only baseline at every operating point (Table~\ref{tab:headline}). At the per-task matched-rate ops the shared encoder is more than an order of magnitude faster than AVIF default-speed and $5$--$8{\times}$ faster than AVIF max-speed across all three tasks, and exceeds $250$ MPx/s for $n{\le}9$---sufficient for 1080p 30 fps over Wi-Fi after accounting for sensor-side concurrency (Fig.~\ref{fig:main_results}), with the SEAOTTER sandwich adding no encode-time overhead. Within-family encode differences (entries marked $^{*}$ in Table~\ref{tab:headline}) are measurement noise, not a real spread between the SEAOTTER-ZS and SEAOTTER-FT encoders.

\runin{Downstream consumer decoding throughput}
The deployed steady-state consumer-side decode of a SEAOTTER artifact is a vanilla JPEG decode followed by $\mathcal{F}^{-1}$. Measured end-to-end on CPU at $384^2$ (Table~\ref{tab:headline}, Appendix~\ref{app:throughput}), SEAOTTER's consumer cost is therefore \textbf{less than a third of AVIF's decoding cost} ($\sim 3.4{\times}$ faster) and \textbf{$100{\times}$ faster than the same FRAPPE codec without the transcode}. The advantage is structural: any consumer decodes the on-disk JPEG with ubiquitous standard hardware and may skip $\mathcal{F}^{-1}$ altogether.

\runin{Deployment-tier suitability}
We check whether each pipeline-and-op cell simultaneously clears three deployment-tier thresholds: BLE ($\text{CR}{\ge}288$, encode $\ge12$ MPx/s), 5G ($\text{CR}{\ge}133$, encode $\ge28$ MPx/s), and Wi-Fi ($\text{CR}{\ge}60$, encode $\ge62$ MPx/s). \mbox{SEAOTTER-FT} clears all three tiers at $n\in\{3,6,9\}$ and the 5G and Wi-Fi tiers at $n{=}12$ (missing only the BLE-tier CR by a thin margin); AVIF clears no tier at any quality we evaluate, and among the neural codecs WaLLoC clears only the BLE and 5G tiers at $p{=}4$ (Table~\ref{tab:deployment_tier}, Appendix~\ref{app:throughput}).

\runin{Why does the transcode help?}
The fine-tune deliberately drives reconstruction PSNR down (from $25.08$ dB in vanilla FRAPPE to $10.39$ dB in \mbox{SEAOTTER-FT} at $n{=}12$; Appendix~\ref{app:rd_tables}) in exchange for downstream accuracy after the transcode. We hypothesize distribution calibration: the softsign companding and DCT-domain $Q^{(k)}$ matrices push the consumer's input back toward the standard JPEG distribution its JPEG-pretrained backbone expects, even where vanilla FRAPPE's outputs look unlike any JPEG image.


\section{Conclusion}
\label{sec:conclusion}

We presented SEAOTTER, a compression framework for cloud robotics that pairs a sensor-embedded autoencoder with a one-time cloud-side transcode into a standards-compliant JPEG file. Across global, dense, and zero-shot perception, the transcode \emph{increases} downstream accuracy over the same DNN-based autoencoder used without it, while producing on-disk artifacts that virtually any data consumer can use.

\runin{Limitations and future work} (i) \emph{Modality coverage.} We test only RGB; depth, IR, multispectral, and hyperspectral signals are a natural extension the framework handles without architectural changes but that we have not characterized. (ii) \emph{Component ablations.} We do not isolate the contributions of the softsign companding, the DCT-domain $Q^{(k)}$ matrices, and the $3{\times}3$ wrapper filter. (iii) \emph{Sensor / lighting variation.} How the learned (approximately YCgCo) color transform varies across sensors, lighting, and lens distortions---and whether per-domain $\mathcal{F}$ pairs help---is left to future work. (iv) \emph{Human perception.} We have not evaluated human perception (e.g., teleoperation) of SEAOTTER-JPEG artifacts versus standard JPEG/AVIF at matched storage rate---important given the nonstandard color space.


\clearpage


\bibliography{refs}

\clearpage


\appendix
\section{Supplementary results and methodology}
\label{app:supplementary}

\subsection{Multi-axis performance summary}
\label{app:summary}

\begin{figure}[H]
\centering
\includegraphics[width=0.7\textwidth]{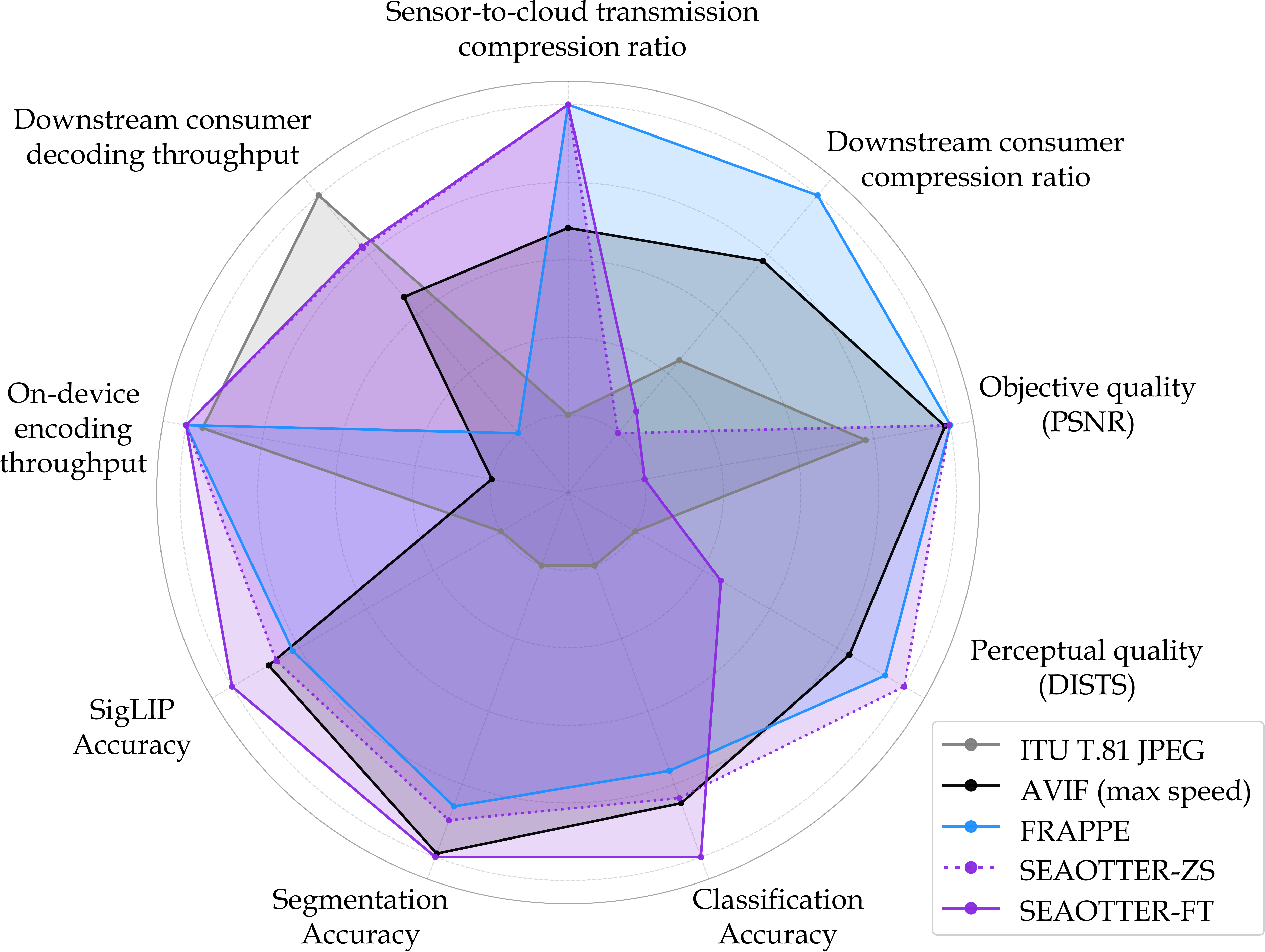}
\caption{Performance trade-offs of SEAOTTER variants vs other codecs.}
\label{fig:radar}
\end{figure}

Figure~\ref{fig:radar} condenses the per-axis results in this appendix into a single view, comparing the SEAOTTER variants against the conventional and neural codec baselines across the sensor-, cloud-, and consumer-side cost axes together with reconstruction quality and downstream accuracy.

\subsection{Learned quantization matrices}
\label{app:quant}

\begin{figure}[!t]
\centering
\includegraphics[width=\textwidth]{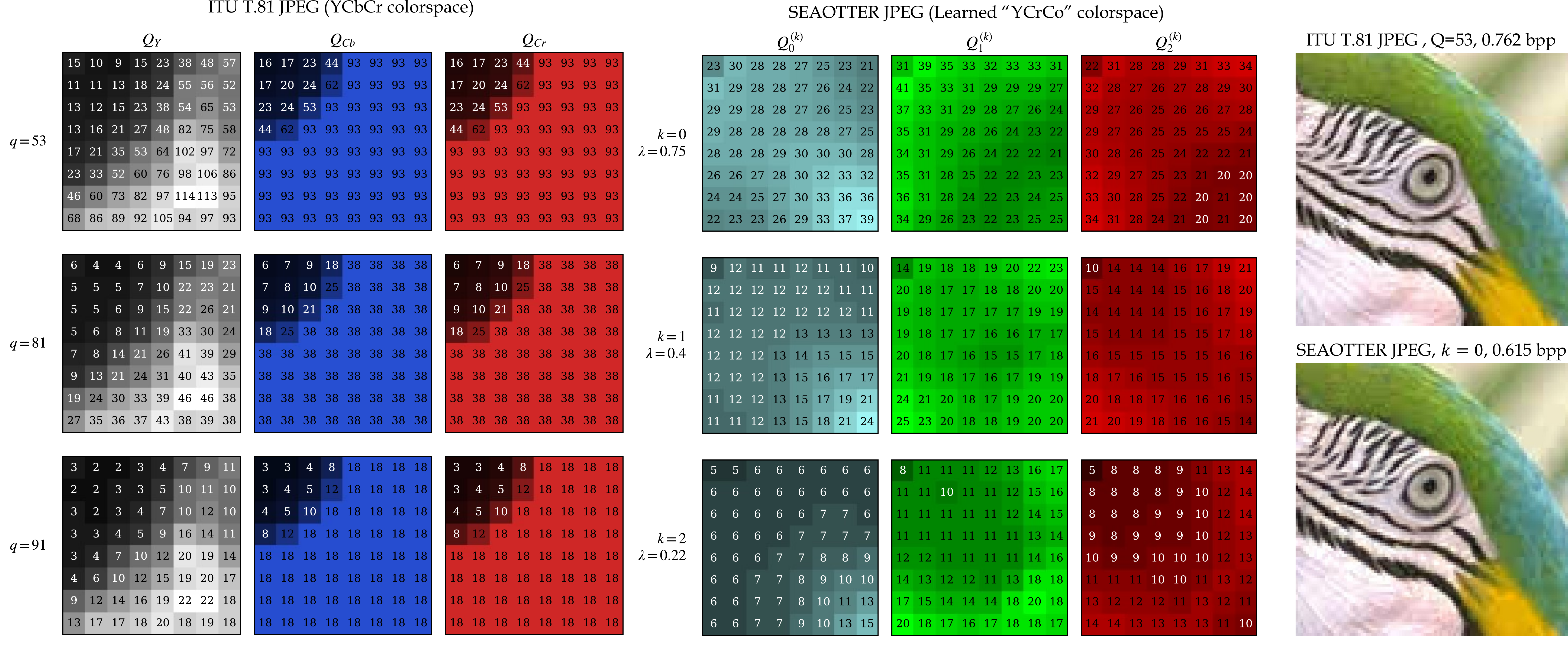}
\caption{Learned per-rate DCT-domain quantization matrices $Q^{(k)}$ for $k=0,1,2$ (top row) alongside the matched-bpp ITU T.81 4:4:4 quantization tables (bottom row). Per-channel colormap hues are derived from $\mathcal{F}$'s learned RGB-mixing kernel; the resulting color space coincides with YCgCo up to per-channel sign.}
\label{fig:quant}
\end{figure}

Figure~\ref{fig:quant} visualizes the three learned DCT-domain quantization matrices $Q^{(k)}$ alongside the matched-bpp ITU T.81 4:4:4 quantization tables. The per-channel colormap hues are derived from $\mathcal{F}$'s learned RGB-mixing kernel and reveal that the learned color space is essentially YCgCo up to per-channel sign: the lowest-bpp matrix is free to crush mid-frequency chroma while the highest-bpp matrix preserves it.

\subsection{Standalone learned JPEG vs ITU T.81 on Kodak}
\label{app:codec_kodak}

\begin{figure}[!t]
\centering
\includegraphics[width=\textwidth]{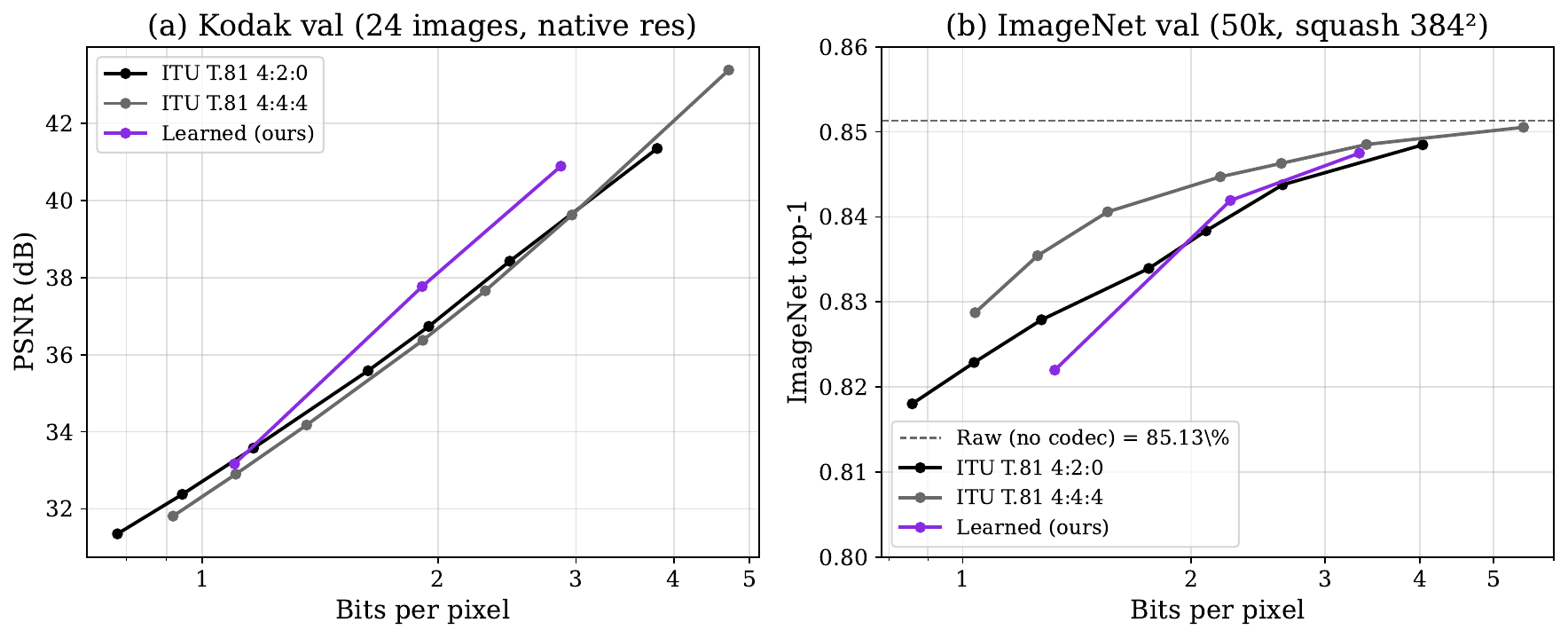}
\caption{Standalone learned JPEG codec versus ITU T.81 (with and without chroma subsampling) on the Kodak validation set at native resolution. SEAOTTER's three trained operating points ($k\in\{0,1,2\}$) dominate matched-bpp ITU T.81 4:4:4 by $+0.27$ / $+1.40$ / $+1.27$ dB in PSNR.}
\label{fig:codec_kodak}
\end{figure}

To isolate the contribution of the learned JPEG sandwich without any FRAPPE-side encoding, we evaluate the trained $(\mathcal{F}, \mathcal{F}^{-1}, Q^{(0)}, Q^{(1)}, Q^{(2)})$ bundle as a standalone codec on the Kodak validation set ($24$ images at native resolution: $16$ images $768{\times}512$, $8$ images $512{\times}768$; no resize, no crop) and compare against ITU T.81 with and without chroma subsampling. The $7$-step quality ladder for the ITU baselines is anchored to the three SEAOTTER operating points by choosing the smallest integer JPEG-$\texttt{sub}{=}0$ quality at which SEAOTTER strictly dominates ITU T.81 4:4:4 on both Kodak PSNR and Kodak bpp, then interpolating intermediate q values. The learned sandwich strictly dominates ITU T.81 4:4:4 in PSNR at all three trained operating points, with margins of $+0.27$ dB / $+1.40$ dB / $+1.27$ dB at matched bpp (Fig.~\ref{fig:codec_kodak} and Table~\ref{tab:codec_kodak}). The standalone-codec evaluation establishes that the sandwich's accuracy gain in the main paper is grounded in a learned representation that is also distortion-favourable in its own right, not a downstream-only artifact of the task-aware fine-tune.

\begin{table}[!t]
\centering
\small
\caption{Standalone codec eval on the Kodak validation set (24 images, native resolution, no FRAPPE upstream).}
\label{tab:codec_kodak}
\begin{tabular}{llrrrrr}
\toprule
Codec & Setting & bpp & PSNR (dB) & SSIM & LPIPS (dB) & DISTS (dB) \\
\midrule
ITU T.81 4:2:0 & $q=39$ & 0.779 & 31.35 & 0.952 & 5.93 & 13.15 \\
ITU T.81 4:2:0 & $q=53$ & 0.943 & 32.38 & 0.963 & 6.68 & 14.56 \\
ITU T.81 4:2:0 & $q=67$ & 1.162 & 33.58 & 0.972 & 7.54 & 16.07 \\
ITU T.81 4:2:0 & $q=81$ & 1.628 & 35.58 & 0.982 & 9.11 & 18.67 \\
ITU T.81 4:2:0 & $q=86$ & 1.946 & 36.73 & 0.985 & 10.05 & 20.12 \\
ITU T.81 4:2:0 & $q=91$ & 2.469 & 38.42 & 0.989 & 11.51 & 21.71 \\
ITU T.81 4:2:0 & $q=96$ & 3.810 & 41.35 & 0.993 & 14.60 & 23.94 \\
\midrule
ITU T.81 4:4:4 & $q=39$ & 0.917 & 31.81 & 0.958 & 6.30 & 15.16 \\
ITU T.81 4:4:4 & $q=53$ & 1.103 & 32.90 & 0.969 & 7.14 & 16.90 \\
ITU T.81 4:4:4 & $q=67$ & 1.359 & 34.17 & 0.978 & 8.08 & 18.65 \\
ITU T.81 4:4:4 & $q=81$ & 1.912 & 36.37 & 0.987 & 9.85 & 22.08 \\
ITU T.81 4:4:4 & $q=86$ & 2.298 & 37.66 & 0.991 & 10.90 & 23.97 \\
ITU T.81 4:4:4 & $q=91$ & 2.965 & 39.62 & 0.994 & 12.60 & 25.83 \\
ITU T.81 4:4:4 & $q=96$ & 4.701 & 43.38 & 0.997 & 16.36 & 28.39 \\
\midrule
SEAOTTER (ours) & $k=0$ & 1.099 & 33.17 & 0.943 & 6.00 & 12.44 \\
SEAOTTER (ours) & $k=1$ & 1.909 & 37.77 & 0.980 & 9.42 & 17.55 \\
SEAOTTER (ours) & $k=2$ & 2.870 & 40.89 & 0.991 & 12.85 & 21.61 \\
\bottomrule
\end{tabular}

\end{table}

\subsection{Standalone learned JPEG on ImageNet}
\label{app:codec_kodak_cls}

Table~\ref{tab:codec_kodak_cls} re-evaluates the same $17$ standalone-codec cells on ImageNet val ($50{,}000$ images, squash-$384^2$ preprocessing, the same teacher as the main classification task) and reports top-1 accuracy plus PSNR. The standalone sandwich is evaluated without the FRAPPE-side upstream that the main paper's SEAOTTER pipeline uses.

\begin{table}[!t]
\centering
\small
\caption{Standalone codec eval on ImageNet val ($50{,}000$ images, squash-$384^2$, \texttt{convnext\_tiny.in12k\_ft\_in1k\_384} teacher). The bpp denominator is pinned at $384^2$. The no-codec ceiling is $85.13\%$ top-1.}
\label{tab:codec_kodak_cls}
\begin{tabular}{llrrr}
\toprule
Codec & Setting & bpp & Top-1 (\%) & PSNR (dB) \\
\midrule
ITU T.81 4:2:0 & $q=39$ & 0.858 & 81.80 & 30.39 \\
ITU T.81 4:2:0 & $q=53$ & 1.036 & 82.29 & 31.29 \\
ITU T.81 4:2:0 & $q=67$ & 1.270 & 82.79 & 32.35 \\
ITU T.81 4:2:0 & $q=81$ & 1.758 & 83.39 & 34.09 \\
ITU T.81 4:2:0 & $q=86$ & 2.091 & 83.84 & 35.06 \\
ITU T.81 4:2:0 & $q=91$ & 2.638 & 84.38 & 36.42 \\
ITU T.81 4:2:0 & $q=96$ & 4.034 & 84.85 & 38.62 \\
\midrule
ITU T.81 4:4:4 & $q=39$ & 1.039 & 82.87 & 31.12 \\
ITU T.81 4:4:4 & $q=53$ & 1.255 & 83.54 & 32.14 \\
ITU T.81 4:4:4 & $q=67$ & 1.552 & 84.06 & 33.38 \\
ITU T.81 4:4:4 & $q=81$ & 2.184 & 84.47 & 35.50 \\
ITU T.81 4:4:4 & $q=86$ & 2.627 & 84.63 & 36.74 \\
ITU T.81 4:4:4 & $q=91$ & 3.401 & 84.85 & 38.65 \\
ITU T.81 4:4:4 & $q=96$ & 5.473 & 85.05 & 42.61 \\
\midrule
SEAOTTER (ours) & $k=0$ & 1.322 & 82.20 & 32.50 \\
SEAOTTER (ours) & $k=1$ & 2.252 & 84.19 & 36.56 \\
SEAOTTER (ours) & $k=2$ & 3.328 & 84.75 & 39.07 \\
\bottomrule
\end{tabular}

\end{table}

\subsection{Per-task rate-distortion details}
\label{app:rd_tables}

Tables~\ref{tab:rd_cls}--\ref{tab:rd_clip} report per-pipeline per-op detail for the three downstream tasks (cls / seg / clip). Each table reports both the transmit bpp (the sensor-uplink rate) and the storage bpp (the on-disk JPEG file after the cloud-side transcode); for codecs without a transcode step the two values coincide. The raw row is the no-codec ceiling for context. \mbox{SEAOTTER-FT}'s reconstruction PSNR is intentionally low because the fine-tune trades pixel fidelity for downstream accuracy (Section~\ref{sec:eval}); we report PSNR for transparency, not as a quality target.

\begin{table}[!h]
\centering
\small
\caption{ImageNet classification (cls): per-cell transmit/storage bpp, top-1 accuracy, and reconstruction PSNR. ImageNet val ($50{,}000$), squash-$384^2$, \texttt{convnext\_tiny.in12k\_ft\_in1k\_384} teacher.}
\label{tab:rd_cls}
\begin{tabular}{llrrrr}
\toprule
Pipeline & Op & Transmit bpp & Storage bpp & Top-1 (\%) & PSNR (dB) \\
\midrule
AVIF & $q=1$ & 0.1458 & 0.1458 & 61.15 & 25.01 \\
 & $q=5$ & 0.1525 & 0.1525 & 62.70 & 25.24 \\
 & $q=6$ & 0.1628 & 0.1628 & 64.67 & 25.56 \\
 & $q=10$ & 0.1840 & 0.1840 & 68.11 & 26.17 \\
 & $q=25$ & 0.2756 & 0.2756 & 75.53 & 27.95 \\
 & $q=50$ & 0.7100 & 0.7100 & 82.09 & 32.28 \\
\midrule
AVIF (max-speed) & $q=1$ & 0.1558 & 0.1558 & 61.02 & 24.65 \\
 & $q=3$ & 0.1624 & 0.1624 & 62.80 & 24.85 \\
 & $q=5$ & 0.1624 & 0.1624 & 62.80 & 24.85 \\
 & $q=10$ & 0.1948 & 0.1948 & 68.22 & 25.70 \\
 & $q=25$ & 0.2934 & 0.2934 & 74.87 & 27.46 \\
 & $q=50$ & 0.7730 & 0.7730 & 82.01 & 31.75 \\
\midrule
FRAPPE & $n=3$ & 0.0122 & 0.0122 & 6.77 & 19.85 \\
 & $n=6$ & 0.0380 & 0.0380 & 26.70 & 22.34 \\
 & $n=9$ & 0.0637 & 0.0637 & 43.21 & 23.71 \\
 & $n=12$ & 0.1086 & 0.1086 & 56.22 & 25.08 \\
 & $n=15$ & 0.3439 & 0.3439 & 73.91 & 28.29 \\
\midrule
WaLLoC & $p=4$ & 0.0428 & 0.0428 & 28.56 & 21.74 \\
 & $p=10.5$ & 0.0971 & 0.0971 & 51.40 & 23.84 \\
 & $p=16$ & 0.1437 & 0.1437 & 60.98 & 24.88 \\
 & $p=36$ & 0.2598 & 0.2598 & 71.70 & 26.57 \\
 & $p=80$ & 0.5312 & 0.5312 & 79.10 & 28.97 \\
 & $p=100$ & 0.6773 & 0.6773 & 80.40 & 30.01 \\
\midrule
SEAOTTER-ZS & $n=3$ & 0.0122 & 0.6441 & 8.30 & 19.84 \\
 & $n=6$ & 0.0380 & 0.8829 & 30.81 & 22.32 \\
 & $n=9$ & 0.0637 & 1.0510 & 47.78 & 23.69 \\
 & $n=12$ & 0.1086 & 1.2822 & 60.25 & 25.04 \\
 & $n=15$ & 0.3439 & 1.7950 & 76.43 & 28.16 \\
\midrule
WaLLoC-SEAOTTER-ZS & $p=4$ & 0.0428 & 0.9038 & 32.42 & 21.73 \\
 & $p=16$ & 0.1437 & 1.1884 & 64.80 & 24.84 \\
 & $p=36$ & 0.2598 & 1.4300 & 73.76 & 26.49 \\
 & $p=80$ & 0.5312 & 1.7925 & 79.41 & 28.79 \\
 & $p=100$ & 0.6773 & 2.0772 & 80.59 & 29.79 \\
\midrule
SEAOTTER-FT & $n=3$ & 0.0122 & 1.6163 & 17.34 & 12.18 \\
 & $n=6$ & 0.0380 & 1.2243 & 46.55 & 12.25 \\
 & $n=9$ & 0.0637 & 1.1352 & 59.69 & 11.64 \\
 & $n=12$ & 0.1086 & 0.9046 & 69.02 & 10.39 \\
 & $n=15$ & 0.3439 & 0.8073 & 77.43 & 9.83 \\
\midrule
Raw (no codec) & -- & 14.6658 & 14.6658 & 85.13 & 120.00 \\
\bottomrule
\end{tabular}

\end{table}

\begin{table}[!h]
\centering
\small
\caption{ADE20K segmentation (seg): per-cell transmit/storage bpp, mIoU, and reconstruction PSNR. ADE20K val ($2{,}000$), squash-$512^2$, UperNet-ConvNeXt-Tiny teacher.}
\label{tab:rd_seg}
\begin{tabular}{llrrrr}
\toprule
Pipeline & Op & Transmit bpp & Storage bpp & mIoU (\%) & PSNR (dB) \\
\midrule
AVIF & $q=1$ & 0.0794 & 0.0794 & 30.85 & 26.67 \\
 & $q=5$ & 0.0860 & 0.0860 & 32.75 & 26.97 \\
 & $q=6$ & 0.0945 & 0.0945 & 33.43 & 27.36 \\
 & $q=10$ & 0.1117 & 0.1117 & 35.27 & 28.09 \\
 & $q=25$ & 0.1850 & 0.1850 & 39.46 & 30.28 \\
 & $q=50$ & 0.5103 & 0.5103 & 43.37 & 35.65 \\
\midrule
AVIF (max-speed) & $q=1$ & 0.0851 & 0.0851 & 30.06 & 26.29 \\
 & $q=3$ & 0.0918 & 0.0918 & 31.47 & 26.56 \\
 & $q=5$ & 0.0918 & 0.0918 & 31.47 & 26.56 \\
 & $q=6$ & 0.1007 & 0.1007 & 32.51 & 26.93 \\
 & $q=7$ & 0.1007 & 0.1007 & 32.51 & 26.93 \\
 & $q=8$ & 0.1106 & 0.1106 & 33.56 & 27.31 \\
 & $q=9$ & 0.1106 & 0.1106 & 33.56 & 27.31 \\
 & $q=10$ & 0.1197 & 0.1197 & 34.05 & 27.63 \\
 & $q=25$ & 0.2021 & 0.2021 & 38.99 & 29.89 \\
 & $q=50$ & 0.5648 & 0.5648 & 43.49 & 35.12 \\
\midrule
FRAPPE & $n=3$ & 0.0096 & 0.0096 & 4.95 & 20.70 \\
 & $n=6$ & 0.0312 & 0.0312 & 17.91 & 23.52 \\
 & $n=9$ & 0.0550 & 0.0550 & 25.24 & 25.14 \\
 & $n=12$ & 0.0939 & 0.0939 & 29.09 & 26.81 \\
 & $n=15$ & 0.3042 & 0.3042 & 38.38 & 30.48 \\
\midrule
WaLLoC & $p=4$ & 0.0331 & 0.0331 & 16.07 & 22.64 \\
 & $p=10.5$ & 0.0817 & 0.0817 & 28.72 & 25.13 \\
 & $p=11$ & 0.0968 & 0.0968 & 30.51 & 25.64 \\
 & $p=12$ & 0.0968 & 0.0968 & 30.51 & 25.64 \\
 & $p=13$ & 0.1127 & 0.1127 & 32.21 & 26.13 \\
 & $p=14$ & 0.1127 & 0.1127 & 32.21 & 26.13 \\
 & $p=15$ & 0.1127 & 0.1127 & 32.21 & 26.13 \\
 & $p=16$ & 0.1300 & 0.1300 & 33.31 & 26.58 \\
 & $p=36$ & 0.2531 & 0.2531 & 38.33 & 28.98 \\
 & $p=80$ & 0.5306 & 0.5306 & 42.12 & 32.07 \\
 & $p=100$ & 0.6320 & 0.6320 & 42.60 & 32.96 \\
\midrule
SEAOTTER-ZS & $n=3$ & 0.0096 & 0.5928 & 4.84 & 20.69 \\
 & $n=6$ & 0.0312 & 0.8188 & 18.15 & 23.51 \\
 & $n=9$ & 0.0550 & 0.9773 & 25.71 & 25.12 \\
 & $n=12$ & 0.0939 & 1.1759 & 30.09 & 26.77 \\
 & $n=15$ & 0.3042 & 1.5222 & 39.42 & 30.37 \\
\midrule
WaLLoC-SEAOTTER-ZS & $p=4$ & 0.0331 & 0.7921 & 16.28 & 22.63 \\
 & $p=16$ & 0.1300 & 1.0685 & 33.58 & 26.54 \\
 & $p=36$ & 0.2531 & 1.3098 & 38.66 & 28.88 \\
 & $p=80$ & 0.5306 & 1.6084 & 42.16 & 31.85 \\
 & $p=100$ & 0.6320 & 1.7689 & 42.65 & 32.74 \\
\midrule
SEAOTTER-FT & $n=3$ & 0.0096 & 0.4575 & 6.66 & 10.13 \\
 & $n=6$ & 0.0312 & 0.4352 & 21.10 & 10.55 \\
 & $n=9$ & 0.0550 & 0.4726 & 28.20 & 10.78 \\
 & $n=12$ & 0.0939 & 0.5240 & 32.77 & 10.76 \\
 & $n=15$ & 0.3042 & 0.6830 & 38.59 & 12.21 \\
\midrule
Raw (no codec) & -- & 11.1280 & 11.1280 & 44.51 & 120.00 \\
\bottomrule
\end{tabular}

\end{table}

\begin{table}[!h]
\centering
\small
\caption{SigLIP-2 zero-shot classification (clip): per-cell transmit/storage bpp, zero-shot top-1, and reconstruction PSNR. ImageNet val ($50{,}000$), naflex preprocessing ($\texttt{max\_num\_patches}{=}256$, $\texttt{patch\_size}{=}16$, $\texttt{snap}{=}32$), SigLIP-2 base-patch16-naflex teacher.}
\label{tab:rd_clip}
\begin{tabular}{llrrrr}
\toprule
Pipeline & Op & Transmit bpp & Storage bpp & Zero-shot Top-1 (\%) & PSNR (dB) \\
\midrule
AVIF & $q=1$ & 0.2503 & 0.2503 & 42.59 & 24.34 \\
 & $q=5$ & 0.2581 & 0.2581 & 44.84 & 24.57 \\
 & $q=6$ & 0.2701 & 0.2701 & 47.98 & 24.91 \\
 & $q=10$ & 0.2941 & 0.2941 & 52.17 & 25.52 \\
 & $q=25$ & 0.3963 & 0.3963 & 61.08 & 27.32 \\
 & $q=50$ & 0.8698 & 0.8698 & 68.67 & 31.66 \\
\midrule
AVIF (max-speed) & $q=1$ & 0.2633 & 0.2633 & 44.19 & 23.96 \\
 & $q=3$ & 0.2709 & 0.2709 & 46.24 & 24.16 \\
 & $q=5$ & 0.2709 & 0.2709 & 46.24 & 24.16 \\
 & $q=10$ & 0.3073 & 0.3073 & 53.29 & 25.02 \\
 & $q=25$ & 0.4169 & 0.4169 & 61.67 & 26.78 \\
 & $q=50$ & 0.9389 & 0.9389 & 68.80 & 31.05 \\
\midrule
FRAPPE & $n=3$ & 0.0187 & 0.0187 & 4.84 & 18.91 \\
 & $n=6$ & 0.0537 & 0.0537 & 16.94 & 21.50 \\
 & $n=9$ & 0.0831 & 0.0831 & 29.03 & 22.92 \\
 & $n=12$ & 0.1417 & 0.1417 & 41.51 & 24.37 \\
 & $n=15$ & 0.4095 & 0.4095 & 60.13 & 27.68 \\
\midrule
WaLLoC & $p=4$ & 0.0511 & 0.0511 & 18.23 & 20.30 \\
 & $p=10.5$ & 0.1101 & 0.1101 & 36.95 & 22.72 \\
 & $p=16$ & 0.1508 & 0.1508 & 44.05 & 23.66 \\
 & $p=36$ & 0.2973 & 0.2973 & 55.85 & 25.81 \\
 & $p=80$ & 0.6195 & 0.6195 & 63.62 & 28.58 \\
 & $p=100$ & 0.7422 & 0.7422 & 64.24 & 29.48 \\
\midrule
SEAOTTER-ZS & $n=3$ & 0.0187 & 0.7441 & 4.84 & 18.90 \\
 & $n=6$ & 0.0537 & 1.0148 & 17.87 & 21.49 \\
 & $n=9$ & 0.0831 & 1.2040 & 30.36 & 22.90 \\
 & $n=12$ & 0.1417 & 1.4650 & 43.34 & 24.32 \\
 & $n=15$ & 0.4095 & 2.0182 & 61.00 & 27.54 \\
\midrule
WaLLoC-SEAOTTER-ZS & $p=4$ & 0.0511 & 1.1057 & 18.37 & 20.30 \\
 & $p=16$ & 0.1508 & 1.3863 & 44.85 & 23.64 \\
 & $p=36$ & 0.2973 & 1.6776 & 56.65 & 25.75 \\
 & $p=80$ & 0.6195 & 2.1065 & 64.14 & 28.39 \\
 & $p=100$ & 0.7422 & 2.3531 & 64.78 & 29.26 \\
\midrule
SEAOTTER-FT & $n=3$ & 0.0187 & 0.4261 & 2.65 & 9.47 \\
 & $n=6$ & 0.0537 & 0.5235 & 20.01 & 12.68 \\
 & $n=9$ & 0.0831 & 0.5879 & 34.82 & 13.43 \\
 & $n=12$ & 0.1417 & 0.6451 & 48.22 & 13.07 \\
 & $n=15$ & 0.4095 & 0.7980 & 61.34 & 12.78 \\
\midrule
Raw (no codec) & -- & 15.1991 & 15.1991 & 69.59 & 120.00 \\
\bottomrule
\end{tabular}

\end{table}

\subsection{Storage-rate trade-offs}
\label{app:storage_rd}

\begin{figure}[!t]
\centering
\includegraphics[width=0.9\textwidth]{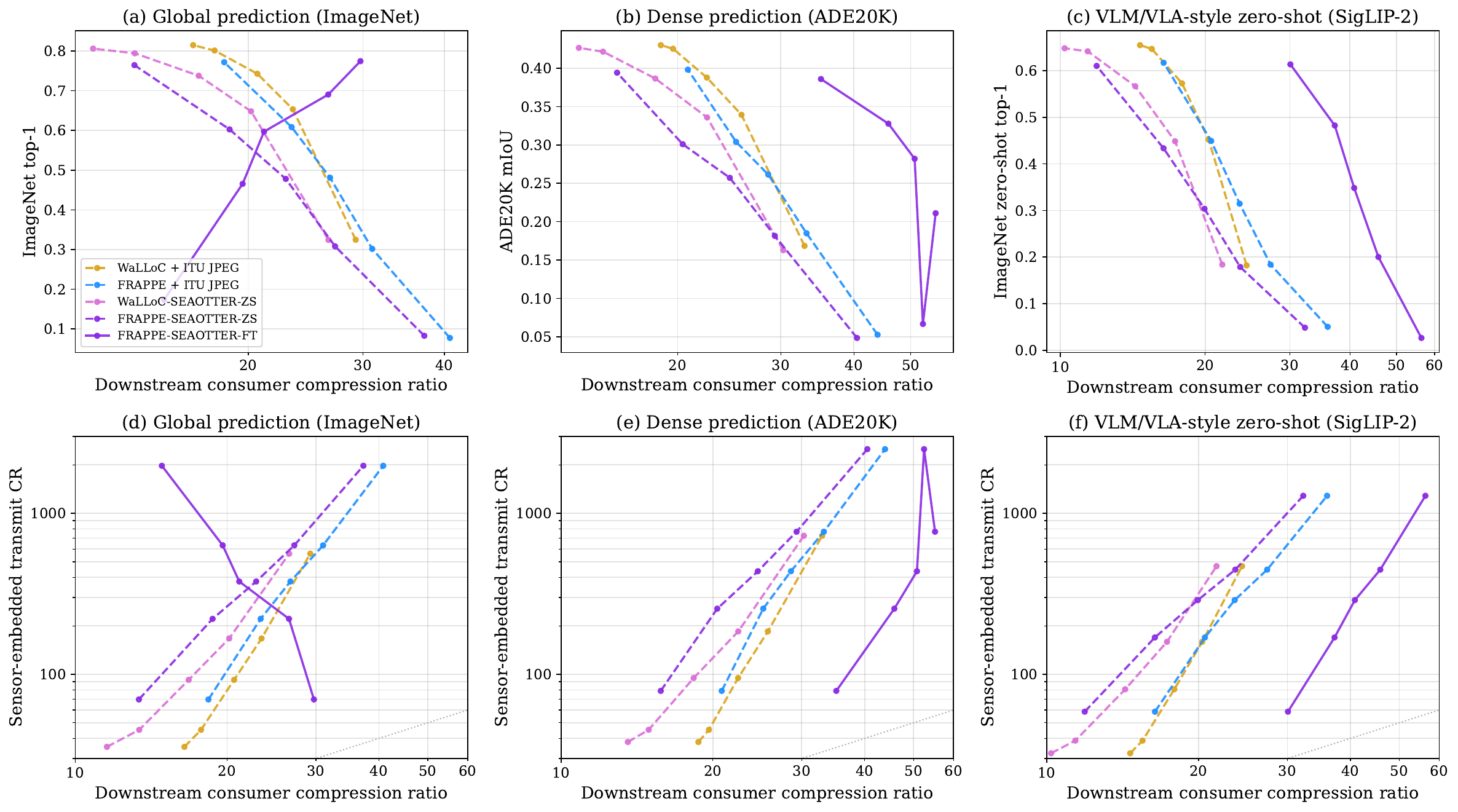}
\caption{Top row (a--c): downstream task accuracy vs.\ the \emph{storage} compression ratio (on-disk JPEG size after transcode). Bottom row (d--f): sensor-embedded transmit CR vs.\ downstream consumer CR, with a $y{=}x$ reference line.}
\label{fig:rd_storage}
\end{figure}

Figure~\ref{fig:rd_storage} re-plots downstream accuracy against the consumer-side \emph{storage} compression ratio (top row) and contrasts the sensor-side transmit CR with the downstream consumer CR (bottom row).

\subsection{Throughput details}
\label{app:throughput}

Table~\ref{tab:throughput} reports wall-clock measurements for sensor-side encode and steady-state consumer-side codec decode (no downstream teacher forward) per pipeline-and-op for the cls task. Each row reports the median over a $32$-image distribution at batch size~$1$ on an AMD EPYC~9354 CPU. For SEAOTTER-family pipelines (\mbox{SEAOTTER-ZS} / \mbox{SEAOTTER-FT} / their WaLLoC-side counterparts) the decode column is the deployed consumer cost: a vanilla JPEG decode followed by the $3{\times}3$ inverse-conv $\mathcal{F}^{-1}$ and per-channel companding. The one-time cloud-side transcode (FRAPPE/WaLLoC neural decode $\to$ sandwich forward $\to$ JPEG encode) is paid once per image and \emph{not} counted in the decode column; it has the same wall-clock as the corresponding FRAPPE-only / WaLLoC-only row's decode.

\begin{table}[!h]
\centering
\small
\caption{Sensor-side encode and steady-state consumer-side codec decode median wall-clock per pipeline-and-op for the cls task. $32$-image distribution at batch size~$1$ on an AMD EPYC~9354 CPU; downstream teacher forward not included (it is the same constant offset across all pipelines). Encode and decode MPx/s are the cls-protocol $384^2$ frame size divided by the corresponding medians. Encode entries marked $^{*}$ all use the same frozen FRAPPE encoder, so the across-row differences in those entries are measurement noise rather than a real spread.}
\label{tab:throughput}
\resizebox{\textwidth}{!}{\begin{tabular}{llrrrrr}
\toprule
Pipeline & Op & Transmit bpp & Enc.\ (ms) & Dec.\ (ms) & Enc.\ (MPx/s) & Dec.\ (MPx/s) \\
\midrule
\shortstack[l]{AVIF\\(default)} & $q=1$ & 0.1458 & 26.75 & 7.47 & 5.51 & 19.75 \\
 & $q=5$ & 0.1525 & 28.40 & 7.45 & 5.19 & 19.79 \\
 & $q=6$ & 0.1628 & 28.67 & 7.42 & 5.14 & 19.87 \\
 & $q=10$ & 0.1840 & 31.99 & 7.56 & 4.61 & 19.50 \\
 & $q=25$ & 0.2756 & 38.09 & 7.72 & 3.87 & 19.10 \\
 & $q=50$ & 0.7100 & 52.63 & 7.92 & 2.80 & 18.62 \\
\midrule
\shortstack[l]{AVIF\\(max speed)} & $q=1$ & 0.1558 & 5.73 & 7.55 & 25.73 & 19.53 \\
 & $q=3$ & 0.1624 & 5.75 & 7.55 & 25.66 & 19.54 \\
 & $q=5$ & 0.1624 & 5.77 & 8.06 & 25.55 & 18.29 \\
 & $q=10$ & 0.1948 & 5.91 & 7.97 & 24.97 & 18.49 \\
 & $q=25$ & 0.2934 & 6.42 & 8.24 & 22.96 & 17.90 \\
 & $q=50$ & 0.7730 & 8.12 & 8.05 & 18.15 & 18.31 \\
\midrule
FRAPPE & $n=3$ & 0.0122 & 0.25 & 223.72 & 601.47$^{*}$ & 0.66 \\
 & $n=6$ & 0.0380 & 0.46 & 236.76 & 317.23$^{*}$ & 0.62 \\
 & $n=9$ & 0.0637 & 0.54 & 212.46 & 271.81$^{*}$ & 0.69 \\
 & $n=12$ & 0.1086 & 0.83 & 218.06 & 177.76$^{*}$ & 0.68 \\
 & $n=15$ & 0.3439 & 1.36 & 214.20 & 108.17$^{*}$ & 0.69 \\
\midrule
WaLLoC & $p=4$ & 0.0428 & 2.59 & 113.77 & 57.00$^{*}$ & 1.30 \\
 & $p=10.5$ & 0.0971 & 3.76 & 45.20 & 39.26$^{*}$ & 3.26 \\
 & $p=16$ & 0.1437 & 4.89 & 37.43 & 30.17$^{*}$ & 3.94 \\
 & $p=36$ & 0.2598 & 7.82 & 67.26 & 18.85$^{*}$ & 2.19 \\
 & $p=80$ & 0.5312 & 13.37 & 131.60 & 11.03$^{*}$ & 1.12 \\
 & $p=100$ & 0.6773 & 13.17 & 135.77 & 11.20$^{*}$ & 1.09 \\
\midrule
SEAOTTER-ZS & $n=3$ & 0.0122 & 0.25 & 2.14 & 601.47$^{*}$ & 68.82 \\
 & $n=6$ & 0.0380 & 0.46 & 2.24 & 317.23$^{*}$ & 65.72 \\
 & $n=9$ & 0.0637 & 0.54 & 2.77 & 271.81$^{*}$ & 53.14 \\
 & $n=12$ & 0.1086 & 0.83 & 2.26 & 177.76$^{*}$ & 65.35 \\
 & $n=15$ & 0.3439 & 1.36 & 2.71 & 108.17$^{*}$ & 54.49 \\
\midrule
\shortstack[l]{SEAOTTER-ZS\\(WaLLoC encode)} & $p=4$ & 0.0428 & 2.59 & 3.18 & 57.00$^{*}$ & 46.39 \\
 & $p=16$ & 0.1437 & 4.89 & 2.92 & 30.17$^{*}$ & 50.58 \\
 & $p=36$ & 0.2598 & 7.82 & 2.90 & 18.85$^{*}$ & 50.80 \\
 & $p=80$ & 0.5312 & 13.37 & 2.21 & 11.03$^{*}$ & 66.71 \\
 & $p=100$ & 0.6773 & 13.17 & 3.08 & 11.20$^{*}$ & 47.88 \\
\midrule
SEAOTTER-FT & $n=3$ & 0.0122 & 0.25 & 2.19 & 601.47$^{*}$ & 67.30 \\
 & $n=6$ & 0.0380 & 0.46 & 2.32 & 317.23$^{*}$ & 63.59 \\
 & $n=9$ & 0.0637 & 0.54 & 2.51 & 271.81$^{*}$ & 58.83 \\
 & $n=12$ & 0.1086 & 0.83 & 2.17 & 177.76$^{*}$ & 67.97 \\
 & $n=15$ & 0.3439 & 1.36 & 2.34 & 108.17$^{*}$ & 62.94 \\
\bottomrule
\end{tabular}
}
\end{table}

\paragraph{Conventional-codec configuration.} All conventional-codec timings use \textbf{Pillow~12.2}: JPEG is encoded and decoded through libjpeg-turbo, and AVIF through Pillow's built-in libavif (libaom for encoding, dav1d for decoding); no \texttt{pillow-avif-plugin} and no GPU or hardware-codec acceleration are involved. Quality is set by Pillow's \texttt{quality} parameter, and the AVIF max-speed variant adds \texttt{speed=10}. AVIF uses libavif's default $4{:}2{:}0$ chroma subsampling; the main-table JPEG likewise uses $4{:}2{:}0$, while the standalone-codec comparison (Appendix~\ref{app:codec_kodak}) uses $4{:}4{:}4$ (\texttt{subsampling=0}). All measurements run in a single Python process at batch size~$1$ on one AMD EPYC~9354 CPU; the harness requests no explicit multi-threading, so each backend runs at its library default with SIMD enabled. Neural encoders run under \texttt{torch.inference\_mode()}. Following the FRAPPE reference harness, each stage is timed with $\texttt{n\_warmup}{=}1$ and $\texttt{n\_measurement}{=}5$ (median per stage), reproducing the reference to within ${\sim}1$--$3\%$.

Table~\ref{tab:deployment_tier} reports, for every pipeline-and-op cell in the comparison, whether the cell simultaneously clears each of the three deployment-tier thresholds defined in Section~\ref{sec:eval}. SEAOTTER and FRAPPE clear all three tiers at low-bitrate operating points; AVIF clears no tier at any quality we evaluate.

\begin{table}[!h]
\centering
\small
\caption{Deployment-tier suitability for every pipeline-and-op cell sorted by descending transmit CR. A pipeline-and-op cell clears a tier iff both its compression ratio and its sensor-side encode throughput exceed the tier threshold. SEAOTTER and FRAPPE clear all three tiers at low-bitrate operating points; AVIF clears no tier at any quality we evaluate.}
\label{tab:deployment_tier}
\begin{tabular}{llrrccc}
\toprule
 &  & Transmit & Encode & \multicolumn{3}{c}{Deployment tier} \\
\cmidrule(lr){5-7}
Pipeline & Op & CR & (MPx/s) & BLE & 5G & Wi-Fi \\
\midrule
FRAPPE & $n=3$ & 1972.6:1 & 601.47 & \checkmark & \checkmark & \checkmark \\
SEAOTTER-ZS & $n=3$ & 1972.6:1 & 601.47 & \checkmark & \checkmark & \checkmark \\
SEAOTTER-FT & $n=3$ & 1972.6:1 & 601.47 & \checkmark & \checkmark & \checkmark \\
FRAPPE & $n=6$ & 632.2:1 & 317.23 & \checkmark & \checkmark & \checkmark \\
SEAOTTER-ZS & $n=6$ & 632.2:1 & 317.23 & \checkmark & \checkmark & \checkmark \\
SEAOTTER-FT & $n=6$ & 632.2:1 & 317.23 & \checkmark & \checkmark & \checkmark \\
WaLLoC & $p=4$ & 561.2:1 & 57.00 & \checkmark & \checkmark & $\cdot$ \\
WaLLoC-SEAOTTER-ZS & $p=4$ & 561.2:1 & 57.00 & \checkmark & \checkmark & $\cdot$ \\
FRAPPE & $n=9$ & 376.9:1 & 271.81 & \checkmark & \checkmark & \checkmark \\
SEAOTTER-ZS & $n=9$ & 376.9:1 & 271.81 & \checkmark & \checkmark & \checkmark \\
SEAOTTER-FT & $n=9$ & 376.9:1 & 271.81 & \checkmark & \checkmark & \checkmark \\
WaLLoC & $p=10.5$ & 247.1:1 & 39.26 & $\cdot$ & \checkmark & $\cdot$ \\
FRAPPE & $n=12$ & 221.0:1 & 177.76 & $\cdot$ & \checkmark & \checkmark \\
SEAOTTER-ZS & $n=12$ & 221.0:1 & 177.76 & $\cdot$ & \checkmark & \checkmark \\
SEAOTTER-FT & $n=12$ & 221.0:1 & 177.76 & $\cdot$ & \checkmark & \checkmark \\
WaLLoC & $p=16$ & 167.0:1 & 30.17 & $\cdot$ & \checkmark & $\cdot$ \\
WaLLoC-SEAOTTER-ZS & $p=16$ & 167.0:1 & 30.17 & $\cdot$ & \checkmark & $\cdot$ \\
AVIF & $q=1$ & 164.6:1 & 5.51 & $\cdot$ & $\cdot$ & $\cdot$ \\
AVIF & $q=5$ & 157.4:1 & 5.19 & $\cdot$ & $\cdot$ & $\cdot$ \\
AVIF (max-speed) & $q=1$ & 154.1:1 & 25.73 & $\cdot$ & $\cdot$ & $\cdot$ \\
AVIF (max-speed) & $q=3$ & 147.7:1 & 25.66 & $\cdot$ & $\cdot$ & $\cdot$ \\
AVIF (max-speed) & $q=5$ & 147.7:1 & 25.55 & $\cdot$ & $\cdot$ & $\cdot$ \\
AVIF & $q=6$ & 147.4:1 & 5.14 & $\cdot$ & $\cdot$ & $\cdot$ \\
AVIF & $q=10$ & 130.4:1 & 4.61 & $\cdot$ & $\cdot$ & $\cdot$ \\
AVIF (max-speed) & $q=10$ & 123.2:1 & 24.97 & $\cdot$ & $\cdot$ & $\cdot$ \\
WaLLoC & $p=36$ & 92.4:1 & 18.85 & $\cdot$ & $\cdot$ & $\cdot$ \\
WaLLoC-SEAOTTER-ZS & $p=36$ & 92.4:1 & 18.85 & $\cdot$ & $\cdot$ & $\cdot$ \\
AVIF & $q=25$ & 87.1:1 & 3.87 & $\cdot$ & $\cdot$ & $\cdot$ \\
AVIF (max-speed) & $q=25$ & 81.8:1 & 22.96 & $\cdot$ & $\cdot$ & $\cdot$ \\
FRAPPE & $n=15$ & 69.8:1 & 108.17 & $\cdot$ & $\cdot$ & \checkmark \\
SEAOTTER-ZS & $n=15$ & 69.8:1 & 108.17 & $\cdot$ & $\cdot$ & \checkmark \\
SEAOTTER-FT & $n=15$ & 69.8:1 & 108.17 & $\cdot$ & $\cdot$ & \checkmark \\
WaLLoC & $p=80$ & 45.2:1 & 11.03 & $\cdot$ & $\cdot$ & $\cdot$ \\
WaLLoC-SEAOTTER-ZS & $p=80$ & 45.2:1 & 11.03 & $\cdot$ & $\cdot$ & $\cdot$ \\
WaLLoC & $p=100$ & 35.4:1 & 11.20 & $\cdot$ & $\cdot$ & $\cdot$ \\
WaLLoC-SEAOTTER-ZS & $p=100$ & 35.4:1 & 11.20 & $\cdot$ & $\cdot$ & $\cdot$ \\
AVIF & $q=50$ & 33.8:1 & 2.80 & $\cdot$ & $\cdot$ & $\cdot$ \\
AVIF (max-speed) & $q=50$ & 31.0:1 & 18.15 & $\cdot$ & $\cdot$ & $\cdot$ \\
\bottomrule
\end{tabular}

\end{table}

\subsection{Training recipe}
\label{app:training_recipe}

For our headline $K{=}3$ configuration we use Lagrange multipliers $(\lambda_1, \lambda_2, \lambda_3) = (0.75, 0.40, 0.22)$ and per-rate loss weights $(w_1, w_2, w_3) = (0.3, 0.7, 1.5)$. Training is performed on the LSDIR dataset~\citep{li2023lsdir} at $480^2$ crops for $4$ epochs, using the Adan optimizer~\citep{xie2024adan} (with \texttt{caution=True}), a raised-cosine learning-rate schedule with base learning rate $1.2{\times}10^{-2}$ (the qtable parameter group runs at half the base rate), batch size $4$ per GPU on $4{\times}$ RTX PRO 6000 GPUs, gradient clipping at $5.0$, and seed $0$. The trained $(\mathcal{F}, \mathcal{F}^{-1})$ pair, the three $Q^{(k)}$ matrices, and the three calibrated rate-proxy scalars are bundled together and published as a single artifact loaded by every downstream experiment via a single library call.

\subsection{Experiment details}
\label{app:experiment_details}

The cls teacher checkpoint is \texttt{convnext\_tiny.in12k\_ft\_in1k\_384}; its no-codec ceiling on this teacher and protocol is $85.13\%$ top-1. The seg no-codec ceiling under the squash-$512^2$ protocol is $44.51\%$ mIoU (about $1.5$ pp below the sliding-window paper-protocol number for the same teacher). The clip naflex preprocessing uses $\texttt{max\_num\_patches}{=}256$, $\texttt{patch\_size}{=}16$, $\texttt{snap}{=}32$; its no-codec ceiling is $69.59\%$ zero-shot top-1. All wall-clock timings are measured at batch size $1$ on an AMD EPYC 9354 CPU paired with an RTX PRO 6000 Blackwell Max-Q GPU.

\end{document}